\documentclass[letterpaper,times]{IONconf}
\usepackage[style=apa]{biblatex}
\title{Logistic-aided Huber M-estimator for robust GNSS positioning}
\author{
    Zhengdao~Li, Penggao~Yan, Li-Ta~Hsu, \textit{The~Hong~Kong~Polytechnic~University}%
}

\begin{document}
\maketitle

\section*{biography}
\biography{Zhengdao Li}{is a doctoral student in the Department of Aeronautical and Aviation Engineering at the Hong Kong Polytechnic University, where he works in the Intelligent Positioning and Navigation Laboratory. His research interests include non-Gaussian GNSS error modelling, robust statistical methods for urban positioning, and integrity monitoring for safety-critical navigation applications.}

\biography{Penggao Yan}{is currently a postdoctoral fellow in the Intelligent Positioning and Navigation Lab at the Department of Aeronautical and Aviation Engineering, The Hong Kong Polytechnic University. His research interests include non-Gaussian error modeling, fault detection and integrity monitoring for localization systems, control-aided localization for autonomous vehicles in extreme conditions, and collaborative positioning and integrity monitoring in urban areas.}

\biography{Li-Ta Hsu}{Li-Ta Hsu is an associate professor at Department of Aeronautical and Aviation Engineering of Hong Kong Polytechnic University. He is Limin Endowed Young Scholar in Aerospace Navigation He received the B.S. and Ph.D. degrees in Aeronauticsand Astronautics from National Cheng Kung University, Taiwan, in 2007 and 2013, respectively. He was a Visiting Researcher with the Faculty of Engineering, University College London and Tokyo University of Marine Science and Technology, in 2012 and 2013, respectively. In 2013, he won a Student Paper Award and two Best Presentation Awards from the Institute of Navigation (ION). He was selected as a Japan Society for the Promotion of Sciences Postdoctoral Fellow with the Institute of Industrial Science, The University of Tokyo and worked from 2014 to 2016. He is an Associate Fellow in the Royal Institute of Navigation. Dr. Hsu currently is members of ION and IEEE and serves as a member of the editorial board and reviewer in professional journals related to GNSS.}

\section*{Abstract}
This paper develops a logistic-aided Huber (LAH) M-estimator for robust GNSS positioning under long-tailed, multipath-affected measurement errors. The key idea is to leverage a logistic measurement error assumption and establish a one-to-one approximation between the logistic-based loglikelihood (i.e., quasi-log-cosh) and the Huber kernel by matching their score functions. This yields closed-form tuning rules for the scale and threshold parameters in the Huber estimator, grounded on logistic error statistical properties. We further show that the proposed LAH estimator preserves comparable efficiency and robustness to the connected logistic-based least quasi-log-cosh (LQLC) estimator. Both Monte Carlo simulations with long-tailed measurement errors and a one-hour urban GNSS dataset confirm that the proposed logistic-statistics-based tuning improves positioning accuracy and precision while suppressing large error spikes. Specifically, LAH reduces the 2D RMSE/STD by 28.03\%/38.83\% versus conventional 95\%-efficiency-based Huber tuning in simulation, and reduces the overall 3D RMSE/STD by 4.85\%/16.68\% in real-world experiments while suppressing large positioning error spikes by up to 51\%.

\section{Introduction}
Precise state estimation is a critical requirement of modern robotics and autonomous systems; however, conventional least squares (LS) estimators based on maximized Gaussian likelihood remain highly sensitive to unpredictable data outliers encountered in real-world operations. Consequently, robust maximum-likelihood-type estimators, commonly known as robust M-estimators, have been widely adopted as a fundamental framework to suppress the influence of contaminated sensor measurements \parencite{huber_robust_1981, huber_robust_2009}.\par

In literature, the Huber M-estimator has gained prominent popularity due to the unique hybrid nature of its loss function. By acting as an $L_2$-norm for small residuals and transitioning to an $L_1$-norm for large outliers, the convex Huber loss function primarily brings two advantages: (1) stable convergence to a unique solution regardless of the initialized point; (2) critical balance between efficiency of $L_2$-norm and robustness of $L_1$-norm \parencite{holland_robust_1977, bosse_robust_2016}. In the Global Navigation Satellite System (GNSS) domain, Huber M-estimation has long been utilized to promote error-resilient positioning in challenging environments, where signal reflection or diffraction frequently induces long-tailed errors. The Huber M-estimation method is extensively employed in standalone stable LS-based positionings \parencite{medina_robust_2019a, crespillo_design_2020, chang_hubers_2005} and is pivotal in robust multi-sensor fusion navigation systems, such as Kalman filters \parencite{yang_robust_2019, gaglione_robust_2019, tseng_robust_2017, chang_hubers_2015, crespillo_tightly_2018} and factor graph optimization \parencite{ahmadi_robust_2025, dominguez-tijero_robust_2025}, to prevent outliers from corrupting the integrated solution. \par

Despite its widespread application, the practical performance of the Huber M-estimator is critically dependent on the precise tuning of the scale and threshold parameters in Huber loss function. Specifically, the scale parameter serves as a normalization factor to standardize residuals, while the threshold parameter defines the transition boundary from a quadratic to a linear profile. In GNSS literature, the scale parameter is typically estimated using the median absolute deviation (MAD) of residuals \parencite{medina_robust_2019a}, or more commonly, predicted using signal-dependent variance models that capture elevation angles \parencite{crespillo_design_2020} or signal strength \parencite{borio_hubers_2018}. The threshold parameter is conventionally fixed at $1.345$. This specific value is theoretically derived to guarantee the Huber M-estimator having a 95\% relative efficiency compared with the LS estimator under perfectly Gaussian measurement errors, while maintaining the estimation robustness against long-tailed measurement errors \parencite{huber_robust_1981, holland_robust_1977}. However, these heuristic parameter-tuning strategies may compromise the performance of Huber M-estimation. As for the scale parameter, the MAD cannot normalize the measurement residuals individually; furthermore, signal-dependent variance models assume Gaussian-distributed nominal errors -- a premise frequently violated in real-world GNSS measurements, especially under severe multipath contamination \parencite{gao_methodology_2009, montenbruck_multignss_2018,heng_statistical_2011, hsu_analysis_2017}. Consequently, the determined scale parameter may cause the loss function to misclassify between inliers and outliers, negatively impacting the accuracy of Huber M-estimation. Besides, the empirical 1.345 efficiency-based threshold may not optimally suit the scenario of GNSS positioning, where observations vary accross different satellites and different environments. Although the recent data-driven method proposed by \textcite{ding_learningenhanced_2022} attempts to adaptively predict the Huber threshold, the work lacks the guideline for scale determination and relies on the quality of learning algorithm and pre-trained data. There is still a dearth of standard approach that systematically facilitates the adjustment of both Huber scale and threshold parameters, leaving the Huber M-estimator vulnerable to misclassified outliers and degraded accuracy in complex environments. \par

To bridge this gap, this study proposes to tune both scale and threshold parameters of Huber M-estimation based on the statistical properties of GNSS errors. We leverage the finding that long-tailed GNSS measurement errors are conveniently yet effectively characterized by the logistic distribution \parencite{el-mowafy_fault_2020a, europeanspaceagency_wp8_2024, balakrishnan_handbook_1991}. Mathematically, the negative log-likelihood of a logistic distribution manifests as the quasi-log-cosh (QLC) loss function \parencite{li_improved_2026}. Crucially, this QLC kernel shares an intrinsic geometric similarity with the Huber loss: the log-cosh-type function of the QLC loss behaves in a \enquote{smoothed Huber} shape, exhibiting a approximately quadratic curve in the small-residual region and a asymptotically linear profile for large outliers \parencite{saleh_statistical_2024a}. Motivated by this continuous and smooth $L_2$-$L_1$ composite resemblance, we aim to establish a one-to-one approximation between a Huber loss and a QLC loss, which is grounded in the logistic density. By constructing this connection, we transform the computationally efficient logistic-aided Huber (LAH) M-estimator into a robust surrogate for the logistic-based least QLC (LQLC) estimator \parencite{li_improved_2026}, which effectively embeds multipath-impacted GNSS error statistics into the Huber loss function. We further show that the proposed kernel approximation yields comparable efficiency and robustness between the LQLC estimator and the linked LAH M-estimator. Monte Carlo simulation and real-world urban GNSS experiment confirm that this logistic-statistics-based tuning improves accuracy and precision under contaminated measurements. With simulated long-tailed errors, LAH reduces the 2D RMSE/STD by 28.03\%/38.83\% versus conventional Huber (CH) tuned based on Gaussian-error scaling the theoretical efficiency of 95\%. In the one-hour urban dataset, LAH solution lowers overall 3D RMSE/STD by 4.85\%/16.68\% while suppressing large error spikes by up to 51\%. \par

The contributions of this study are threefold: 
\begin{enumerate}
    \item Proposed standard procedures to adjust both scale and threshold parameters in a Huber M-estimator for robust GNSS positioning (i.e., LAH M-estimation), through a closed-form parameterization mapping that directly embeds the logistic-based prior GNSS error statistics into the tuning constants of the Huber loss.
    \item Rigorously justified the mapping by demonstrating approximated estimation efficiency and robustness between LAH M-estimator and the connected logistic-based LQLC estimator.
    \item Validated the improved estimation performance of the proposed LAH M-estimation compared to CH M-estimation, through both simulation and the real-world experiment.
\end{enumerate}

The remaining parts of this paper are structured as follows. Section \ref{sec: LAH formulation} introduces the formulation of the LAH M-estimator. Section \ref{sec: Experimental results} provides both simulated and real-world GNSS experiments to validate the effectiveness and robustness of the proposed LAH M-estimator in comparison to conventional methods. Section \ref{sec: Conclusion and future work} concludes this study and proposes prospective research directions.

\section{Logistic-aided Huber (LAH) M-estimator} \label{sec: LAH formulation}
In this section, we will start from the basics of logistic-based maximum likelihood estimator (MLE). Hereinafter, by discussing the approximate relationship between the QLC and the Huber kernel functions, we explore the link between logistic error statistics and a Huber loss.

\subsection{Logistic error model and LQLC estimator}

Considering $n$ satellites in view, the linearized GNSS pseudorange measurement model gives
\begin{equation}
    \mathbf{y} = \mathbf{H}\mathbf{x} + \bm{\varepsilon}, \label{equ: pseudo mea model}
\end{equation}
where \textbf{y} is the vector of difference between the measured and predicted pseudorange; $\textbf{H}$ is the geometry matrix; $\textbf{x}$ is the state vector which contains user positions and clock bias; and $\bm{\varepsilon}$ is the vector of measurement error. As suggested by recent GNSS literature \parencite{el-mowafy_fault_2020a, li_improved_2026, europeanspaceagency_wp8_2024}, the long-tailed logistic distribution is leveraged to model real-world measurement errors. It offers a more accurate empirical fit than the traditional Gaussian model, yet retains a comparably straightforward parameterization. Mathematically, a logistic distribution is characterized by a location ($m$) and scale parameter ($s$), with the PDF ($f_L$) and CDF ($F_L$) expressed by:
\begin{align}
    f_L(x; m, s) &= \frac{e^{-\frac{x-m}{s}}}{s \prt{1+e^{-\frac{x-m}{s}}}^2} = \frac{1}{s  \prt{e^{\frac{x-m}{s}}+e^{-\frac{x-m}{s}} +2}}, \label{equ: logi pdf} \\
    F_L(x; c, s) &= \frac{1}{1+e^{-\frac{x-m}{s}}} \label{equ: logi cdf}.
\end{align}
The work of \textcite{li_improved_2026} has shown that the MLE of the states $\mathbf{x}$ based on logistic error model is given by:
\begin{equation} \label{equ: LQLC estimator with raw notation}
    \hat{\textbf{x}} =  \argmin \sum_{i=1}^n \ln \prt{\cosh \prt{\frac{\textbf{y}_{(i)}-\textbf{H}_{(i,:)} \textbf{x}}{s_i}}+1} = \argmin \sum_{i=1}^n J_\text{LQLC}\prt{\bar{r}_i},
\end{equation}
where for the $i$-th measurement, $\bar{r}_i = (\mathbf{y}_{(i)} - \mathbf{H}_{(i,:)}\mathbf{x}) / s_i$ represents the normalized residual and $s_i$ is the associated logistic scale parameter. Here, $\mathbf{y}_{(i)}$ denotes the $i$-th element of the measurement vector; $\mathbf{H}_{(i,:)}$ represents the $i$-th row of the design matrix; the term $J_\text{LQLC}(\cdot)$ denotes the loss function of the least quasi-log-cosh (LQLC) estimator . The notations will be maintained throughout this paper.



\subsection{Approximate relationship between Huber and QLC loss function}
To mathematically build the parameterization connection between the Huber and QLC kernels, we analyze their respective score functions (or noted as $\psi$ functions), which represent the derivative of the loss function and closely relate to the efficiency and robustness of the M-estimators \parencite{hampel_robust_1986}. For the $i^{th}$ measurement residual $r_i$, and Huber loss function gives:
\begin{equation}    \label{equ: huber loss function}
    J_\text{Huber}(r_i; c_i, \sigma_i)= \left\{
        \begin{array}{lr} 
            \frac{1}{2}\prt{\frac{r_i}{\sigma_i}}^2, & \frac{|r_i|}{\sigma_i} \leq c_i \\
            c_i\prt{\frac{|r_i|}{\sigma_i} - \frac{1}{2}c_i}, & \frac{|r_i|}{\sigma_i} > c_i
        \end{array},
    \right.
\end{equation}
where $\sigma_i$ and $c_i$ denote the Huber scale and threshold parameters, respectively. When $\frac{|r_i|}{\sigma_i}\leq c_i$, Huber score function gives:
\begin{equation}  \label{equ: huber psi function case 1}
    \psi_\text{Huber}(r_i; c_i, \sigma_i) = \frac{r_i}{\sigma_i^2}.
\end{equation}
When $\frac{|r_i|}{\sigma_i}> c_i$, the score function is
\begin{equation}  \label{equ: huber psi function case 2}
    \psi_\text{Huber}(r_i; c_i, \sigma_i) = \frac{c_i}{\sigma_i}\cdot \text{sgn}(r_i),
\end{equation}
where sgn represents the sign function. According to Equation \eqref{equ: LQLC estimator with raw notation}, QLC loss function and the corresponding score function regarding residual $r_i$ are given by:
\begin{align}  
    &J_\text{LQLC}(r_i; s_i) = \ln\prt{\cosh\prt{\frac{r_i}{s_i}}+1}, \label{equ: QLC loss function} \\
    &\psi_\text{LQLC}(r_i; s_i) = \frac{1}{s_i}\tanh\prt{\frac{r_i}{2s_i}}. \label{equ: QLC psi function}
\end{align}
When $\frac{|r_i|}{s_i}\rightarrow 0$, we apply the Taylor series for the function $\tanh(x)$ near 0 by
\begin{equation}
    \tanh(x) = x-\frac{x^3}{3}+\frac{2x^5}{15}+O\prt{x^6}.
\end{equation}
Taking the first-order expansion, we further formulate Equation \eqref{equ: QLC psi function} as:
\begin{equation}  \label{equ: QLC psi function case 1}
    \psi_\text{LQLC}(r_i; s_i) \approx \frac{r_i}{2s_i^2}. 
\end{equation}
When $\frac{|r_i|}{s_i}\rightarrow \infty$, Equation \eqref{equ: QLC psi function} gives
\begin{equation}  \label{equ: QLC psi function case 2}
    \psi_\text{LQLC}(r_i; s_i) = \frac{1}{s_i} \cdot \frac{\exp\prt{\frac{|r_i|}{2s_i}} - \exp\prt{-\frac{|r_i|}{2s_i}}}{\exp\prt{\frac{|r_i|}{2s_i}} + \exp\prt{-\frac{|r_i|}{2s_i}}} =  \frac{1}{s_i} \prt{1-\frac{2}{1+\exp\prt{\frac{|r_i|}{s_i}}}} \approx \frac{1}{s_i}\cdot \text{sgn}\prt{r_i}.
\end{equation}

For the purpose of building approximate relationship between LQLC and Huber loss at both small-residual and large-residual regions, we propose the following procedures.\par

When residual $r_i$ is near the origin, we match the score function of Huber and LQLC loss by equating Equations \eqref{equ: huber psi function case 1} and \eqref{equ: QLC psi function case 1}, yielding that
\begin{equation}   \label{equ: logistic-aided approximation a}
    \sigma_i^2 = 2s_i^2.
\end{equation}
When residual $r_i$ is approaching infinity, we likewise equate Equations \eqref{equ: huber psi function case 2} and \eqref{equ: QLC psi function case 2} to obtain that
\begin{equation}    \label{equ: logistic-aided approximation b}
    \frac{c_i}{\sigma_i} = \frac{1}{s_i}.
\end{equation}
By substituting the relationship from Equation \eqref{equ: logistic-aided approximation a} into Equation \eqref{equ: logistic-aided approximation b}, we obtain the mapping between LQLC and Huber loss parameters:
\begin{align} \label{equ: LAH parameter tuning}
    \sigma_i &= \sqrt{2}s_i, \notag \\
    c_i &= \sqrt{2}.
\end{align}

\begin{figure}[H]
    \centering   
    \captionsetup[subfigure]{skip=3pt, margin={2em, 0pt}}
    \addsubFig{0.45}{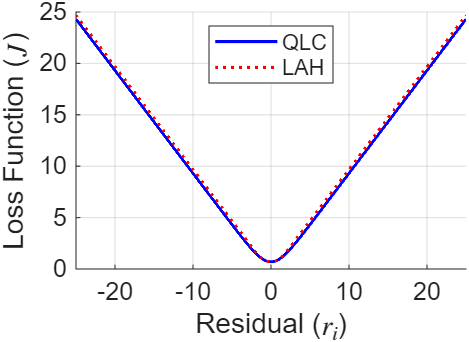}{}
    \addsubFig{0.45}{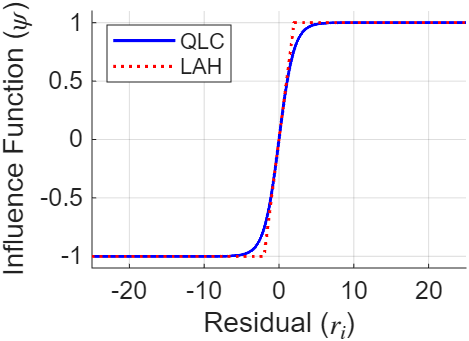}{}\\
    \caption{Comparison between LQLC (with unit scale parameter) and the LAH estimators, in terms of (a) loss function and (b) score function.}
    \label{fig: approximate loss score}
\end{figure}

Using this mapping for all individual loss functions, we introduce the logistic-aided Huber (LAH) M-estimator, where the Huber tuning parameters are strictly determined by the logistic-based statistical properties of the measurement errors. To visualize the relationship between the LQLC and the linked LAH, we set the logistic scale parameter to unity (i.e., $s_i=1$) and compare the two kernels from the perspective of their loss and score functions in Figure \ref{fig: approximate loss score}. The curves demonstrate that the LAH closely adheres to the LQLC at both small and large residuals. Instead of calculating the complex exponential terms required by the QLC loss, the LAH utilizes a piecewise function that bounds the influence of large outliers linearly. This structural simplicity allows the LAH to act as a computationally efficient surrogate for the QLC loss, thereby making the LAH M-estimator a highly practical alternative to the LQLC M-estimator.\par


\subsection{Theoretical validation on the approximation}
In order to justify the proposed approximation between QLC and Huber loss function, we aim to theoretically demonstrate that similar efficiency and robustness exist between the LQLC estimator and the linked LAH M-estimator. To simplify the discussion, we assume on the homoscedaticity case where each measurement error is independent and identically distirbuted for the linear system in Equation \eqref{equ: pseudo mea model}. 

\subsubsection{Efficiency similarity}
\begin{figure}[H]
    \centering   
    \captionsetup[subfigure]{skip=3pt, margin={2em, 0pt}}
    \addsubFig{0.48}{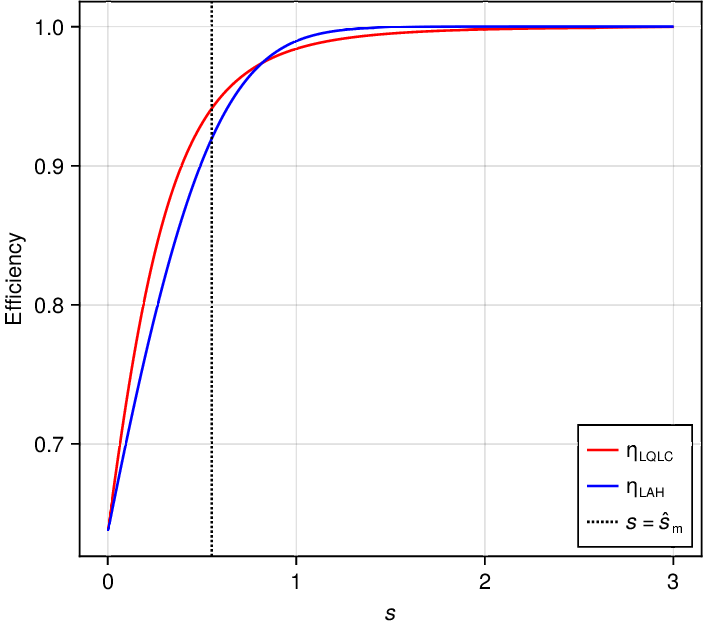}{}
    \addsubFig{0.48}{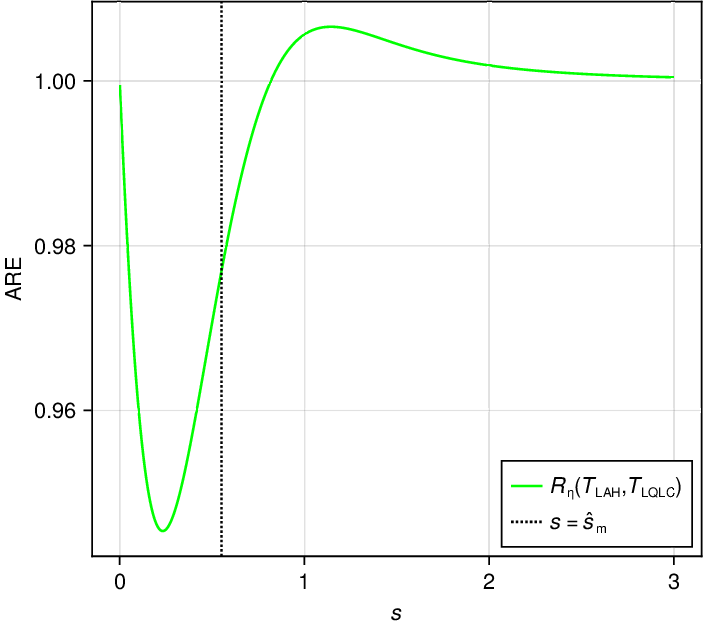}{}\\
    \caption{Efficiency analysis for LQLC and LAH estimators: (a) the change in efficiency of both estimators when the logistic scale parameter $s$ increases from 0.0 to 3.0, and (b) ARE of LQLC estimator compared to LAH estimator.}
    \label{fig: efficiency approximate analysis}
\end{figure}

When using robust M-estimation, a classic concern is the inflated uncertainty compared to the LS estimator under the perfectly Gaussian error assumption. This question introduces the definition of efficiency for a general unbiased M-estimator (noted as $T$) in the linear regression problem when number of measurements approaches infinity:
\begin{equation} \label{equ: def efficiency}
    \eta_T = R_\eta(T, T_\text{LS})=\frac{\det \brk{V(T_\text{LS})}}{\det \brk{V(T)}},
\end{equation}
where $R_\eta(T, T_\text{LS})$ represents the asymptotic relative efficiency (ARE) of estimator $T$ compared to LS estimator $T_\text{LS}$; det means the determinant; and the $V(T)$ denotes asymptotic variance of an estimator $T$, further defined by
\begin{equation}  \label{equ: def AV}
    V(T) = \frac{\text{E}\brk{\psi_{T}^2}}{\prt{\text{E}\brk{\psi'_{T}}}^2}\cdot \prt{\mathbf{H}^\top \mathbf{H}}^{-1},
\end{equation}
where E($\cdot$) stands for the expectation under standard Gaussian assumption and $\psi_{T}$ denotes the $\psi$ function of the M-estimator $T$. The efficiency of LQLC and LAH estimators can be expressed in functions of $s$, respectively, as:
\begin{align}
    \eta_{T_\text{LQLC}}(s) &= \frac{\det \brk{V(T_\text{LS})}}{\det \brk{V(T_\text{LQLC})}} = \frac{\prt{\int_{-\infty}^{\infty} \left[ \frac{1}{2s^2}\text{sech}^2 \left(\frac{r}{2s}\right) \right] f(r) dr}^2}{\int_{-\infty}^{\infty} \left[ \frac{1}{s}\tanh\left(\frac{r}{2s}\right) \right]^2 f(r) dr}. \\
    \eta_{T_\text{LAH}}(s) &= \frac{\det \brk{V(T_\text{LS})}}{\det \brk{V(T_\text{LAH})}} =\frac{\prt{\int_{-2s}^{2s} \frac{1}{2s^2} f(r) dr}^2}{\int_{-2s}^{2s} \left( \frac{r}{2s^2} \right)^2 f(r) dr + 2 \int_{2s}^{\infty} \left( \frac{1}{s} \right)^2 f(r) dr},
\end{align}
where $f(r)$ represents the standard normal PDF. The detailed derivations have been provided in Appendix \ref{app: efficiency calculation}.

Figure \ref{fig: efficiency approximate analysis}a displays how the efficiency of both LQLC estimator and LAH M-estimator change when the magnitude of logistic scale parameter $s$ varies between 0.0 and 3.0. As can be seen, both curves monotonically increases and approaches 1.0 when the $s$ gradually increases. The discrepancy in efficiency of two estimators is reflected in Figure \ref{fig: efficiency approximate analysis}b, where the ARE of LQLC estimator compared to LAH M-estimator is bounded between 0.95 and 1.01. The results reveal the approximated efficiency between estimators when $s$ varies, with a maximum difference below 5\%. \par

Notably, for the standard Gaussian measurement error distribution, we may further leverage a moment estimator to determine the logistic scale parameter $\hat{s}_m$ as: 
\begin{equation}
    \hat{s}_m = \frac{\sqrt{3}}{\pi} \sigma_{Gaussian}  \approx 0.55.
\end{equation}
As depicted in Figure \ref{fig: efficiency approximate analysis}a, when $s=\hat{s}_m$, the LQLC and LAH estimators have efficiency of 94.12\% and 91.95\%, respectively. The results demonstrate the proposed LAH estimator yields a slight efficiency loss of roughly 3\%, compared to the 95\% efficiency requirement in conventional Huber (CH) M-estimator, using the estimated $\hat{s}_m$. Besides, Figure \ref{fig: efficiency approximate analysis}b evidence the close efficiency performance between the two estimators at $s=\hat{s}_m$, with a difference of roughly 2\%. \par





\subsubsection{Robustness similarity}
\begin{figure}[H]
    \centering   
    \captionsetup[subfigure]{skip=3pt, margin={2em, 0pt}}
    \addsubFig{0.48}{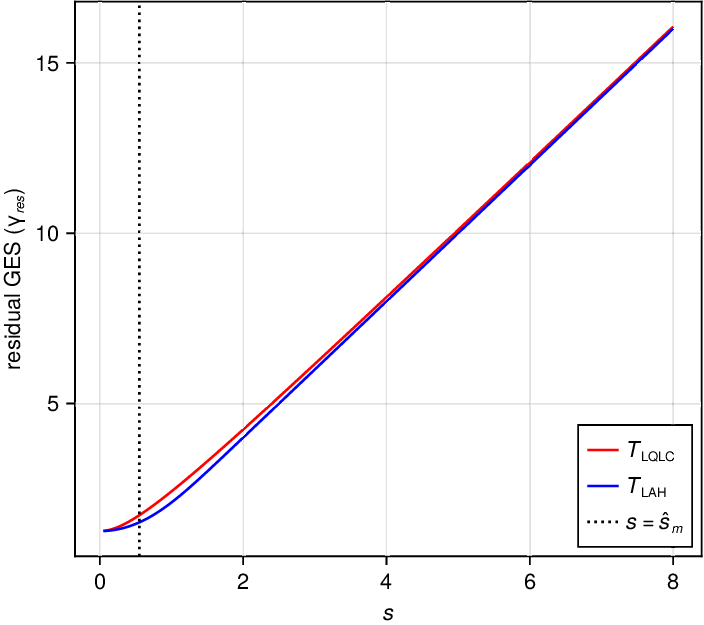}{}
    \addsubFig{0.48}{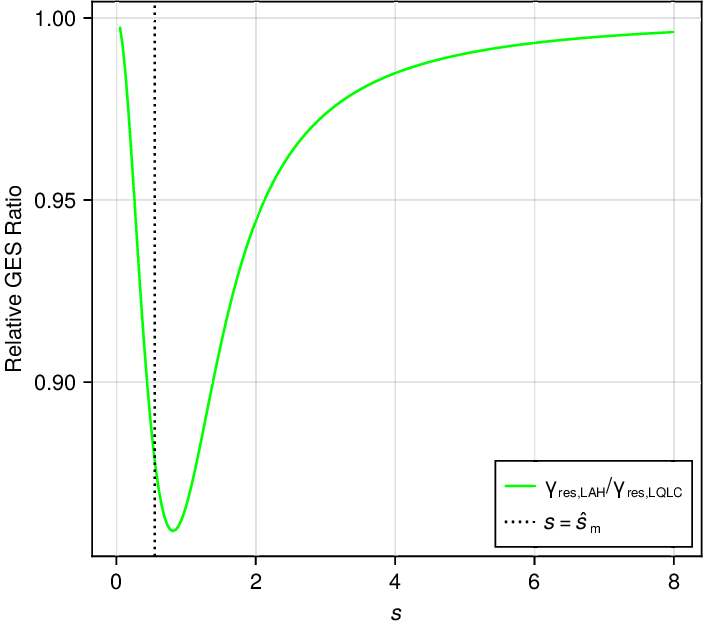}{}\\
    \caption{Robustness analysis for LQLC and LAH estimators: (a) the change in residual GES ($\gamma_{\text{res}}$) of both estimators when the logistic scale parameter $s$ increases from 0.0 to 8.0, and (b) relative GES of LQLC estimator compared to LAH estimator.}
    \label{fig: robustness approximate analysis}
\end{figure}

To mathematically validate that the proposed LAH M-estimator preserves the robust characteristics of the LQLC estimator, we evaluate their respective local and global robustness against measurement anomalies through gross error sensitivity (GES) and measurement breakdown point (BDP), respectively \parencite{hampel_robust_1986}.

In robust statistics, the local robustness of an estimator against severe measurement error is commonly quantified by the GES. A finite GES dictates that no matter how large an individual error becomes, its "pull" on the estimation solution is strictly capped; the magnitude of GES measures the vulnerability of estimator against measurement outliers. In the GNSS context, the satellites's positions are known and the pseudorange measurement equation is linearized using the assumption that initial guess is close to the true states, thus we may assume the geometry matrix $\textbf{H}$ is fixed during the estimation. For a M-estimator with a fixed $\textbf{H}$, the influence of a single outlier is dominated by the loss kernel's score function $\psi$. We introduce the residual GES ($\gamma_{\text{res}}$) to isolate this local robustness:
\begin{equation}
    \gamma_{\text{res}} = \frac{\sup_{r} |\psi(r)|}{\text{E}\left[\psi'(r)\right]},
\end{equation}
where $\text{E}[\cdot]$ denotes the expectation under the nominal Gaussian assumption. The detailed derivation for $\gamma_{\text{res}}$ has been provided in Appendix \ref{app: r GES calculation}.

For the standard LS estimator, the linear score function $\psi_\text{LS}(r) = r$ is unbounded. This yields an infinite residual GES ($\gamma_{\text{res}, \text{LS}} \equiv \infty$), indicating a critical vulnerability where a single massive pseudorange error can introduce unbounded distortions to the estimated states. In contrast, the score functions of robust M-estimators clip the influence of large residuals. Based on the parameter mapping in Equation \eqref{equ: LAH parameter tuning}, the proposed LAH kernel gives the residual GES:
\begin{equation}
    \gamma_{\text{res}, \text{LAH}} =\frac{1}{s\prt{\int_{-\infty}^{\infty} \left[ \frac{1}{2s^2}\text{sech}^2\left(\frac{r}{2s}\right) \right] f(r) dr}}  .
\end{equation}
Besides, the residual GES of LQLC estimator is given by
\begin{equation}
    \gamma_{\text{res}, \text{LQLC}} = \frac{1}{s\prt{\int_{-2s}^{2s} \frac{1}{2s^2} f(r) dr}}  .
\end{equation}
Because of the difference in score derivatives ($\psi'$), the two estimators has slightly diverged expectations, impacting the denominator of the $\gamma_{\text{res}, \text{LAH}}$ and $\gamma_{\text{res}, \text{LQLC}}$.

Figure \ref{fig: robustness approximate analysis}a illustrates the $\gamma_{\text{res}}$ for the estimators as the logistic scale parameter $s$ varies between 0.05 and 6.0. As the assumed scale $s$ increases (representing a score function approximating towards that of LS estimator), outlier clipping of both estimators is weaken, causing the $\gamma_{\text{res}}$ to predictably increase. The relative residual GES between the two estimators is reflected in Figure \ref{fig: robustness approximate analysis}b. A minimum of 0.86 shows that the discrepancy percentage in residual GES is bounded within 14\%. Given the estimated scale parameter $\hat{s}_m \approx 0.55$, the relative GES ratio ($\gamma_{\text{res},\text{LAH}} / \gamma_{\text{res},\text{LQLC}}$) is approximately 0.88. Because a lower GES indicates more effective outlier mitigation, the results demonstrate that the proposed LAH estimator not only mimics the local robustness of the LQLC estimator but actually yields a lower vulnerability for severe gross errors.

Furthermore, the global robustness of the estimator is characterized by its finite-sample BDP. Standard regression estimators are inherently vulnerable to leverage points in the design matrix (or the geometry matrix in GNSS) $\mathbf{H}$ , yielding a theoretical BDP of zero \parencite{maronna_robust_2019}. However, in GNSS positioning, the deterministic $\textbf{H}$ allows us to evaluate global robustness via the measurement BDP (or $y$-BDP). Because both $\psi_\text{QLC}$ and $\psi_\text{LAH}$ are monotone and bounded, they achieve the maximum theoretical measurement BDP of $\varepsilon_y = 0.5$ \parencite{rousseeuw_robust_1987}.

In conclusion, replacing the computationally demanding QLC kernel with the simplified LAH piecewise kernel provides similar and marginally higher local robustness bounds against severe pseudorange anomalies, while preserving the 50\% global measurement outlier tolerance.

\section{Experimental results} \label{sec: Experimental results}
To evaluate the performance of LAH M-estimator, we compare it against the standard LS method and conventional Huber (CH) M-estimator with Gaussian scaling and a $1.345$ threshold, using both simulation and real-world experiment.

\subsection{Monte Carlo simulation}

In the first simulation, we design a 2D localization task for a target sensor, which is configured to receive distance measurements from eight surrounding anchors. The true distances between the sensor and anchors are set equally to $1 \times 10^6$ m, which is large enough to safely omit linearization effects, closely mirroring the scenario of GNSS positioning. To replicate the long-tailed nature of real-world GNSS errors, the true distances from each anchor are deliberately contaminated with random noise generated from a Student's t-distribution with two degrees of freedom. By generating a large-sample set of distance observations for each anchor, we fit the sample data with both Gaussian and logistic distributions. These empirical fits are subsequently used to determine the scale parameters for each individual loss function of the CH and LAH M-estimators, respectively. The position estimation is executed across $1 \times 10^5$ Monte Carlo experimental trials utilizing the iteratively reweighted least squares (IRLS) framework \parencite{holland_robust_1977}.\par

\begin{figure}[H]
    \centering   
    \captionsetup[subfigure]{skip=3pt, margin={2em, 0pt}}
    \addsubFig{0.48}{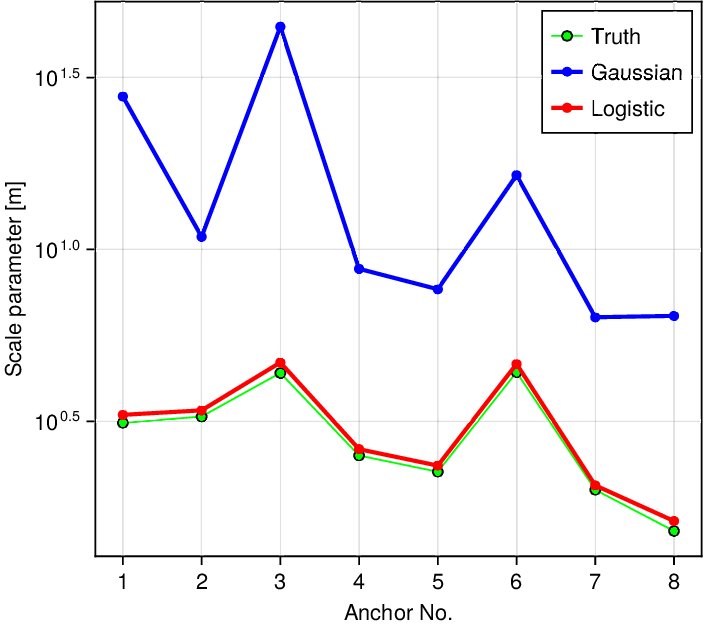}{}   
    \hfill
    \addsubFig{0.48}{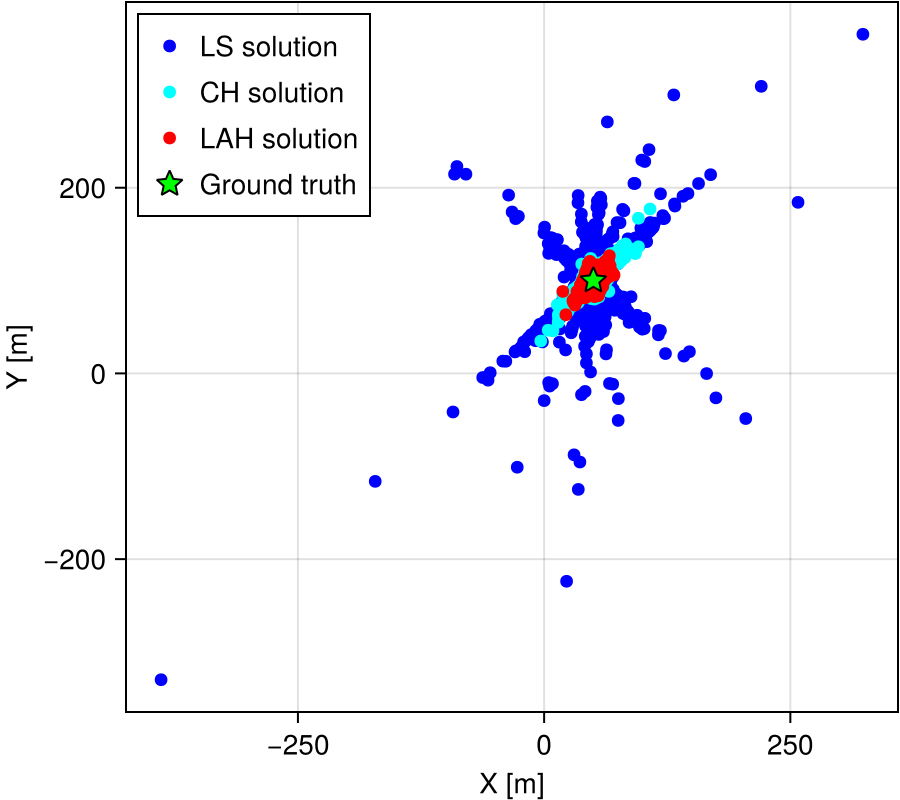}{} 
    \addsubFig{0.48}{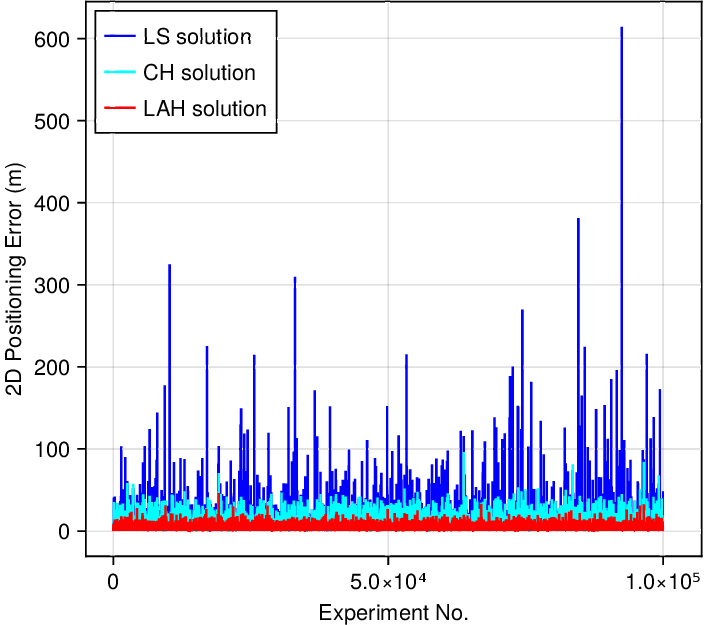}{}
    \hfill
    \addsubFig{0.48}{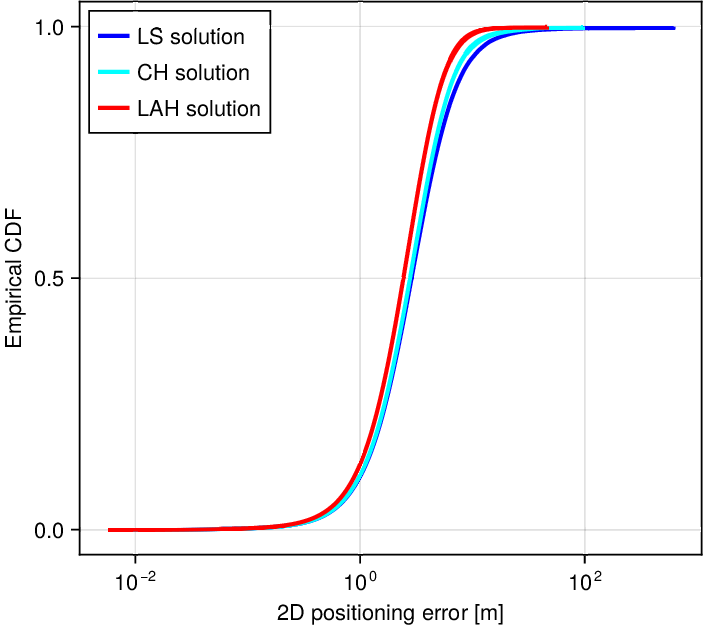}{} 
    \caption{Monte Carlo simulation results: (a) true and maximum-likelihood-estimated scale parameters for measurements from the 8 anchors; (b) 2D positioning in X-Y coordinate; (c) 2D positioning error throughout the $1\times 10^5$ experiments; (d) empirical CDF of 2D positioning errors.}
    \label{fig: exp1 simulation}
\end{figure}

The fundamental advantage of the proposed LAH estimator over the CH baseline is rooted in its scale parameterization, as illustrated in Figure \ref{fig: exp1 simulation}a. Because the Student's t-distribution generates long-tailed measurement errors, fitting a Gaussian model onto the data may inflate the estimated scale parameter (i.e., the Gaussian sigma). Consequently, the Gaussian-derived scale parameter for the CH estimator exhibits high magnitude and volatility across the eight anchors. Especially, the extreme magnitude occurs for scale parameter of anchor No. 1 and 3, where the Gaussian scale roughly tenfolds the truth scale parameter in Student's t-distributions. In contrast, the logistic fit captures the long-tailed behavior considerably more accurately, yielding a stably estimated scale parameter closely matching the truth. Because Huber-based M-estimators rely on this scale parameter to normalize residuals, precise empirical scaling is critical to the estimator's success. By preventing the scale inflation inherent to the Gaussian fit, the LAH estimator ensures that measurement residuals are not inappropriately normalized. This accurate normalization allows the Huber threshold to properly classify severe measurement errors as outliers, correctly applying the linear penalty to suppress their influence. Conversely, the inflated scale in the CH approach inadvertently shrinks the normalized residuals, causing true measurement outliers to be misclassified as inliers and to corrupt the position solution. \par

This theoretical advantage at the measurement level directly translates into superior estimation precision and the suppression of extreme positioning errors during the Monte Carlo experiments. The spatial distribution of the estimated 2D positions is first visualized in Figure \ref{fig: exp1 simulation}b. The unrobustified LS solutions are highly dispersed across the X-Y coordinate plane, reflecting a severe sensitivity to the long-tailed measurement errors. Although the CH solutions (cyan) form a more concentrated cluster, the proposed LAH solutions (red) are the most densely aggregated around the ground truth. This visual clustering confirms that the LAH M-estimation effectively localizes the position solution with superior spatial precision.\par

Figure \ref{fig: exp1 simulation}c depicts the 2D positioning error throughout the $1 \times 10^5$ simulations. The LS method is shown to be extremely vulnerable to the injected measurement noise; the time-series curve experiences severe positioning error spikes that frequently exceed 150 m, occasionally reaching up to 600 m. While the CH M-estimator largely mitigates the most extreme deviations, the inflated scaling factor derived from the Gaussian fit causes the CH solution to still generate moderate positioning errors that plateau around 40 m. In contrast, the proposed LAH M-estimator benefits from the precise logistic-based scaling factor, generating the most stable and consistently lowest 2D positioning errors throughout the simulations. In Figure \ref{fig: exp1 simulation}d, the overall reliability of the proposed method is manifested by the empirical cumulative distribution function (CDF) of the 2D positioning errors. The LAH curve exhibits a noticeable leftward shift compared to both the LS and CH baselines, indicating that the LAH estimator yields a significantly higher probability of achieving a low-magnitude positioning error. Furthermore, the ending abscissa of the CDF curves reflects the magnitude of the maximum positioning error, which coincides directly with the peak error spikes observed in Figure \ref{fig: exp1 simulation}c.\par

Finally, these visual and theoretical advantages of the LAH M-estimator are quantitatively confirmed by the overall error metrics detailed in Table \ref{tab: exp1 simulation}. We evaluate the solution accuracy using the 2D root mean square error (RMSE) and the precision using the 2D standard deviation (STD) of the errors. As shown in Table 1, the LAH estimator reduces the 2D RMSE to 3.60 m (compared to the CH solution of 5.01 m and the LS solution of 7.73 m) and the 2D STD to 2.13 m (compared to 3.49 m for CH and 6.53 m for LS). Notably, the proposed LAH M-estimator achieves a significant performance gain, further reducing the 2D RMSE by 28.03\% and decreasing the 2D STD by 38.83\% relative to the CH M-estimation approach.\par

\begin{table}[h]
    \caption{Overall 2D error metrics in the Monte Carlo simulation.}
    \label{tab: exp1 simulation}
    \begin{tblr}
            {colspec={X[c]X[c]X[c]X[c]X[c]},
            width=\textwidth,
            row{even} = {white,font=\small},
            row{odd} = {bg=black!10,font=\small},
            row{1} = {bg=black!20,font=\bfseries\small},
            hline{Z} = {1pt,solid,black!60},
            rowsep=3pt
            }
            \textbf{Metric}& \textbf{LS solution} & \textbf{CH solution} & \textbf{LAH solution} & \textbf{Percentage reduction (LAH v.s. CH)} \\
            2D RMSE & 7.73m & 5.01m & 3.60m &  28.03\% \\
            2D STD & 6.53m & 3.49m & 2.13m  &  38.83\%
    \end{tblr}
\end{table}

\subsection{Real-world experiment}

Building upon the simulation results, we subsequently evaluate the estimators in a 3D GNSS single point positioning scenario. We utilize a one-hour, 1 Hz dataset collected in the urban environment in Hong Kong using a u-blox F9P receiver tracking GPS L1 signals. Since GNSS error statistics are known to be significantly impacted by satellite elevation \parencite{park_satellite_1996, hu_gpsbdsgalileo_2021}, we develop an elevation-dependent scale model to adaptively generate the scale parameter for individual GNSS observations. To construct this model, we segment the satellite elevation angles into three-degree bins, each of which contains sufficient number of samples. The measurement errors within each bin are fitted with both Gaussian and logistic distributions to extract the corresponding scale parameters. During the estimation process, the elevation-dependent scale model outputs the scale parameters when the satellite observation lies in the corresponding bin of elevation. \par

\begin{figure}[H]
    \centering
    \captionsetup[subfigure]{skip=3pt, margin={2em, 0pt}}
    \addsubFig{0.48}{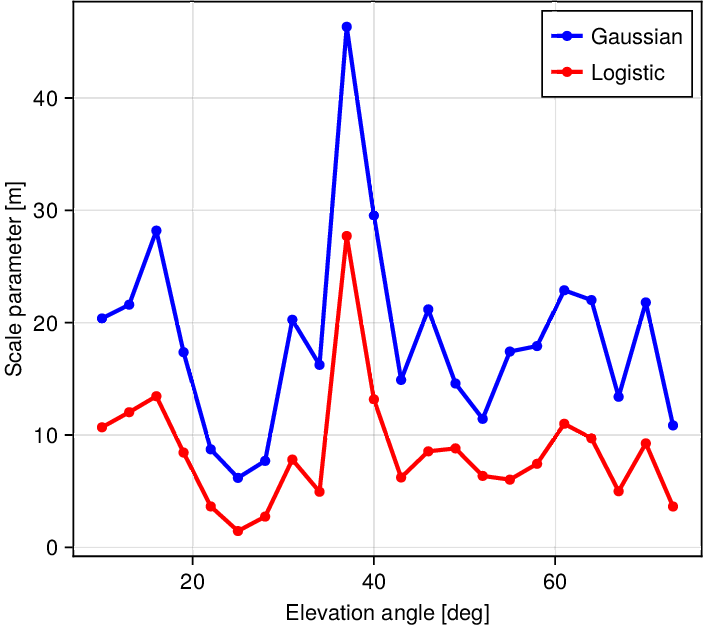}{}   
    \hfill
    \addsubFig{0.48}{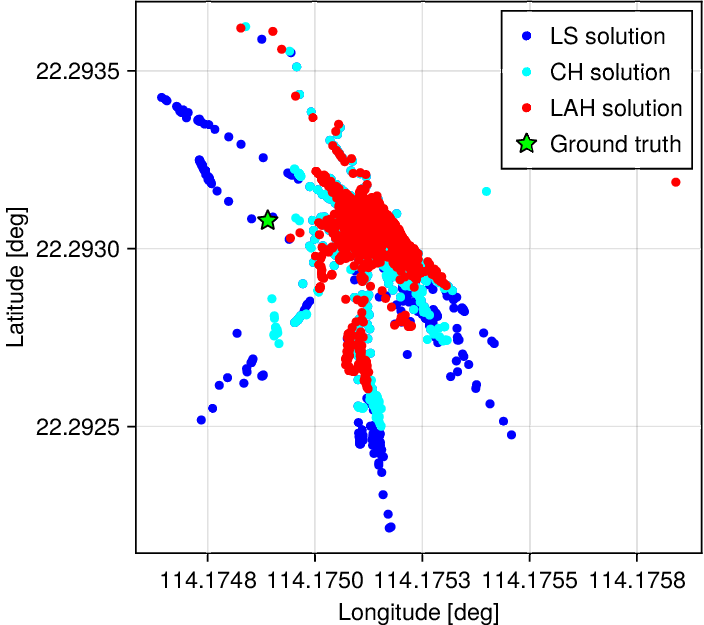}{}
    \addsubFig{0.48}{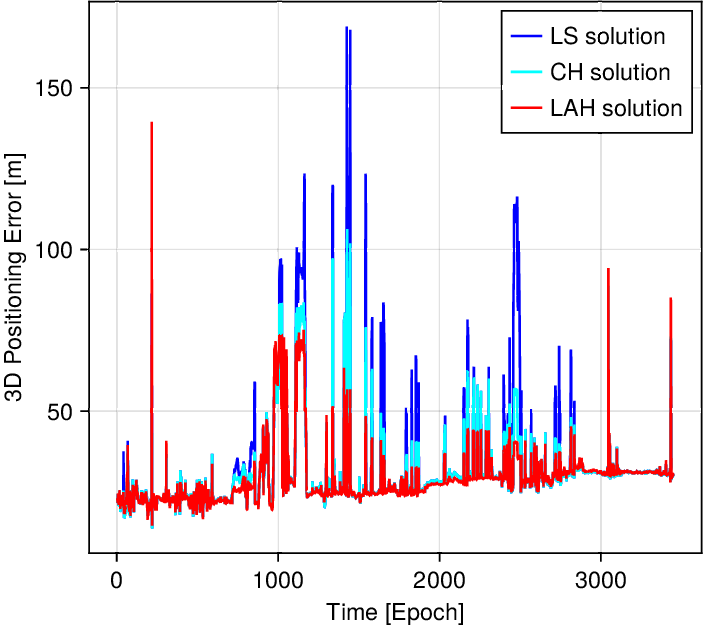}{} 
    \hfill
    \addsubFig{0.48}{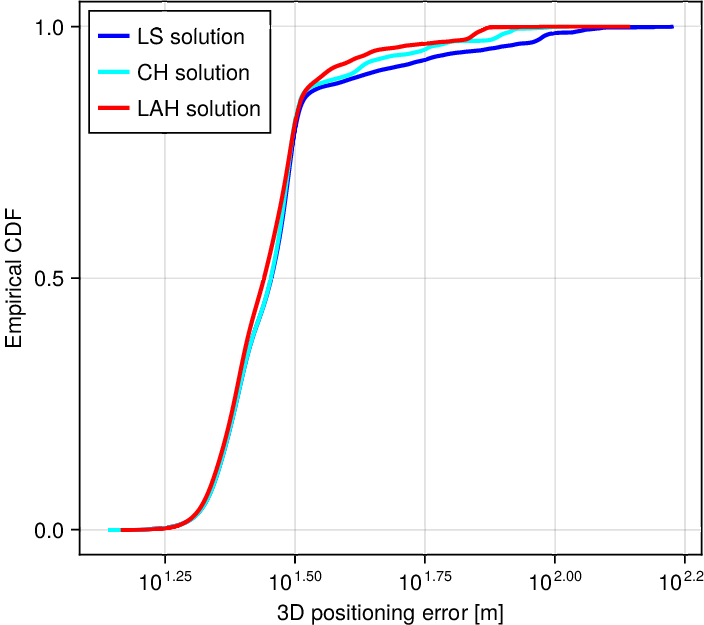}{} 
    \caption{Real-world experiment results: (a) elevation-dependent scale models based on Gaussian and logistic distributions; (b) 2D positioning in longitude-latitude coordinate; (c) 3D positioning error throughout the $1\times 10^5$ experiments; (d) empirical CDF of 3D positioning errors.}
    \label{fig: exp2 realworld}
\end{figure}

The practical necessity of the LAH M-estimator is demonstrated by the elevation-dependent scale analysis in Figure \ref{fig: exp2 realworld}a. In real-world urban environments, GNSS signals frequently suffer from multipath and non-line-of-sight effects, which result in long-tailed measurement errors. Although Figure \ref{fig: exp2 realworld}a does not exhibit the extremely inflated Gaussian scale parameters observed in the simulation experiments, the magnitude of the Gaussian scale parameter consistently exceeds that of the logistic scale parameter. This larger magnitude can induce greater scaling factors within the CH M-estimator compared to the LAH approach, causing the CH loss function to misclassify large measurement outliers as inliers. Conversely, the logistic fit intrinsically captures the long-tailed nature of the measurement errors more accurately, providing the LAH estimator with a smaller and more stable empirical scale to correctly normalize residuals and bound the their influence during estimation.\par

As visualized in the longitude-latitude scatter plot of Figure \ref{fig: exp2 realworld}b, the LS solutions are largely dispersed due to their high sensitivity to extreme measurement errors. In contrast, positioning solutions from the Huber-based robust M-estimator form significantly more concentrated clusters. In particular, the proposed LAH M-estimator yields positions with the highest precision, exhibiting the densest solution cluster. Figure \ref{fig: exp2 realworld}c depicts the 3D positioning error over the 3,451-epoch evaluation. Discrepancies among the three approaches are clearly evidenced between epochs 700 and 2800: the LS solution (blue) generates the highest error spikes, the CH solution (cyan) moderately reduces the extreme errors, and the proposed LAH solution (red) further significantly lowers the error magnitude. Notably, compared to the CH solution, the LAH M-estimator achieves the largest error reduction of 51.21\% at epoch 1429. The improved positioning performance of the LAH M-estimator can be attributed to the fitted logistic scale parameter, which characterizes long-tailed measurement errors more accurately than the Gaussian sigma. Consequently, the logistic-based scaling factor in the LAH loss function more appropriately normalizes the residuals, enabling outliers to be more effectively classified by the threshold and thereby having their influence downweighted. These mechanisms lead to the enhanced estimation performance of the LAH M-estimator. The overall effectiveness of LAH M-estimator on positioning error mitigation is statistically validated by the empirical error CDF in Figure \ref{fig: exp2 realworld}c. The curve of LAH solution exhibits a distinct leftward shift, especially when 3D error exceeds 35m, confirming a consistently higher probability of achieving smaller 3D positioning errors under real-world conditions.\par

Table \ref{tab: exp2 realworld} summarizes the overall metrics quantifying the 3D positioning errors produced by the three estimation approaches. Over the one-hour evaluation period, the LAH estimator reduces the overall 3D RMSE to 30.68 m (compared to 32.24 m for the CH solution and 35.89 m for the LS solution) and the 3D STD to 9.32 m (compared to 11.19 m for CH and 16.42 m for LS). Ultimately, compared to the CH M-estimator, the LAH M-estimator improves both positioning accuracy and precision, yielding a 4.85\% reduction in 3D RMSE and a 16.68\% reduction in 3D STD. \par

These real-world experiment complements the Monte Carlo simulations, demonstrating the consistently superior estimation performance (in terms of both accuracy and precision) achieved by the proposed LAH M-estimator. The results also validate the feasibility of LAH parameter tuning and confirm the expectation that the proposed tuning methods yield improved estimation solutions under long-tailed measurement errors compared to the conventional tuning approach. \par



\begin{table}[ht]
    \caption{Overall 3D error metrics in the real-world experiment.}
    \label{tab: exp2 realworld}
    \begin{tblr}
            {colspec={X[c]X[c]X[c]X[c]X[c]},
            width=\textwidth,
            row{even} = {white,font=\small},
            row{odd} = {bg=black!10,font=\small},
            row{1} = {bg=black!20,font=\bfseries\small},
            hline{Z} = {1pt,solid,black!60},
            rowsep=3pt
            }
            \textbf{Metric}& \textbf{LS solution} & \textbf{CH solution} & \textbf{LAH solution} & \textbf{Percentage reduction (LAH v.s. CH)} \\
            3D RMSE & 35.89m & 32.24m & 30.68m & 4.85\% \\
            3D STD  & 16.42m & 11.19m & 9.32m  & 16.68\% 
    \end{tblr}
\end{table}

\section{Conclusion and future work} \label{sec: Conclusion and future work}
This study promotes the rigorous application of robust Huber M-estimation in the GNSS field by introducing a systematic parameter-tuning workflow. Specifically, we leverage the empirical suitability of the logistic distribution for long-tailed GNSS measurement errors and construct a one-to-one approximation between the logistic-based quasi-log-cosh (QLC) kernel and the Huber kernel via score function matching. This yields closed-form tuning rules for both scale and threshold parameters of the individual logistic-aided Huber (LAH) loss, bridging the statistical properties of GNSS measurement errors with the parameterization of Huber loss function. We also theoretically validate that similar efficiency and robustness exist between the LQLC estimator and the linked LAH M-estimator given by the proposed kernel approximation. Both Monte Carlo simulation and real-world urban GNSS positioning experiments validate that this statistically grounded parameter tuning improves both accuracy and precision under contaminated non-Gaussian measurements. In the simulation with long-tailed errors, LAH M-estimator reduces the 2D RMSE and STD by 28.03\% and 38.83\% compared with the conventional Huber (CH) M-estimator. In the one-hour urban dataset, LAH achieved consistent improvements, reducing the overall 3D RMSE and STD by 4.85\% and 16.68\%, while substantially suppressing large positioning error spikes up to 51\%. \par

Future work will involve extensively validating the performance gains of the LAH estimator over the CH approach across a broader range of GNSS-challenged environments. To further enhance the estimator's adaptive capabilities, the current elevation-dependent logistic scale model could be expanded into a more precise and multivariate weighting scheme, which depends on additional signal quality indicators (e.g., carrier-to-noise density ratio ($C/N_0$)). Furthermore, the LAH estimation can be extended to differential GNSS (DGNSS) or real-time kinematic (RTK) applications that rely on similar linear regression architectures. Beyond standalone positioning, this statistically-driven LAH M-estimation can yield superior resilience and computable uncertainty, possessing significant potential for advanced multi-sensor fusion architectures such as Kalman filters and factor graph optimization.

\printbibliography

\newpage
\appendix
\section*{APPENDIX}
\vspace{1em}

\subsection*{Efficiency calculation for LQLC and LAH estimators}
\label{app: efficiency calculation}

Based on Equation \eqref{equ: def AV}, the asymptotic variance (V) of LS estimator gives:
\begin{equation} \label{aequ: AV LS}
    V(\text{LS}) = \frac{E[\psi_\text{LS}^2]}{\prt{E[\psi_\text{LS}']}^2}\cdot \prt{\mathbf{H}^\top \mathbf{H}}^{-1} = \frac{\text{Var}\brk{r} + \prt{E[r]}^2}{1^2} \cdot \prt{\mathbf{H}^\top \mathbf{H}}^{-1}= \prt{\mathbf{H}^\top \mathbf{H}}^{-1}.
\end{equation}

Similarly, 
\begin{equation} \label{aequ: AV LQLC}
    V(\text{LQLC}) = \frac{E[\psi_\text{LQLC}^2]}{\prt{E[\psi_\text{LQLC}']}^2}\cdot \prt{\mathbf{H}^\top \mathbf{H}}^{-1}, 
\end{equation}
where
\begin{align}
    E[\psi_\text{LQLC}'] &= \int_{-\infty}^{\infty} \left[ \frac{1}{2s^2}\text{sech}^2\left(\frac{r}{2s}\right) \right] f(r) dr, \label{aequ: E LQLC p} \\
    E[\psi_\text{LQLC}^2] &= \int_{-\infty}^{\infty} \left[ \frac{1}{s}\tanh\left(\frac{r}{2s}\right) \right]^2 f(r) dr.
\end{align}


Combining Equation \eqref{equ: huber psi function case 1} and \eqref{equ: huber psi function case 2} with the proposed Huber parameter setting in Equation \eqref{equ: LAH parameter tuning}, we obtain that
\begin{equation}    
    \psi_\text{LAH}(r; s)= \left\{
        \begin{array}{lr} 
            \frac{r}{2s^2},           & |r| \leq 2s \\
            \frac{1}{s}\text{sgn}(r), & |r| > 2s
        \end{array},
    \right.
\end{equation}

\begin{equation}    
    \psi_\text{LAH}'(r; s)= \left\{
        \begin{array}{lr} 
            \frac{1}{2s^2},           & |r| \leq 2s \\
            0,                        & |r| > 2s
        \end{array}.
    \right.
\end{equation}
Therefore, the LAH M-estimator has
\begin{equation} \label{aequ: AV LAH}
    V(\text{LAH}) = \frac{E[\psi_\text{LAH}^2]}{\prt{E[\psi_\text{LAH}']}^2}\cdot \prt{\mathbf{H}^\top \mathbf{H}}^{-1}, 
\end{equation}
where
\begin{align}
    E[\psi_\text{LAH}'] &= \int_{-2s}^{2s} \frac{1}{2s^2} f(r) dr, \label{aequ: E LAH p} \\
    E[\psi_\text{LAH}^2] &= \int_{-2s}^{2s} \left( \frac{r}{2s^2} \right)^2 f(r) dr + 2 \int_{2s}^{\infty} \left( \frac{1}{s} \right)^2 f(r) dr.
\end{align}

Combining Equation \eqref{equ: def efficiency}, \eqref{aequ: AV LS}, and \eqref{aequ: AV LQLC}, we obtain the efficiency of LQLC estimator by:
\begin{equation}
    \eta_{T_\text{LQLC}}(s) = \frac{\det \brk{V(T_\text{LS})}}{\det \brk{V(T_\text{LQLC})}} = \frac{\prt{\int_{-\infty}^{\infty} \left[ \frac{1}{2s^2}\text{sech}^2 \left(\frac{r}{2s}\right) \right] f(r) dr}^2}{\int_{-\infty}^{\infty} \left[ \frac{1}{s}\tanh\left(\frac{r}{2s}\right) \right]^2 f(r) dr}.
\end{equation}

Combining Equation \eqref{equ: def efficiency}, \eqref{aequ: AV LS}, and \eqref{aequ: AV LAH}, we obtain the efficiency of LAH estimator by:
\begin{equation}
    \eta_{T_\text{LAH}}(s) = \frac{\det \brk{V(T_\text{LS})}}{\det \brk{V(T_\text{LAH})}} =\frac{\prt{\int_{-2s}^{2s} \frac{1}{2s^2} f(r) dr}^2}{\int_{-2s}^{2s} \left( \frac{r}{2s^2} \right)^2 f(r) dr + 2 \int_{2s}^{\infty} \left( \frac{1}{s} \right)^2 f(r) dr}.
\end{equation}

\subsection*{Residual GES formulation}
\label{app: r GES calculation}
In robust statistics, the influence function (IF) describes the marginal impact that an infinitesimal contamination at a single data point exerts on the state estimate. For the GNSS linear regression problem in Equation \eqref{equ: pseudo mea model}, we evaluate the physical distortion caused by a single measurement anomaly (e.g., due to severe multipath impact) occurring at index $k \in [1,n]$. According to the work of \textcite{hampel_robust_1986}, the influence function (IF) for a multidimensional M-estimator evaluated at this $k$-th measurement is defined as:
\begin{equation} \label{equ: def IF}
    \text{IF}(\psi, r_k) =  \mathbf{M}^{-1} \prt{\psi(r_k) \mathbf{H}_{(k,:)}^\top},
\end{equation}
where $r_k = \mathbf{y}_{(k)} - \mathbf{H}_{(k,:)}\mathbf{x}$ is the measurement residual, $\psi$ is the score function, and $\mathbf{M}$ is the expected Jacobian matrix evaluated under the nominal uncontaminated error distribution using all observations in the snapshot. Assuming a fixed satellite geometry and independent errors, $\mathbf{M}$ simplifies to:
\begin{equation} \label{equ: def IF M}
    \mathbf{M} = -\text{E} \left[ \frac{\partial \left( \psi(r)\mathbf{H}^\top \right)}{\partial \mathbf{x}} \right] =  \text{E} \brk{ \psi'(r) \mathbf{H}^\top \mathbf{H}}. 
\end{equation}
To evaluate the absolute worst-case vulnerability of the estimator, we utilize the gross error sensitivity (GES, denoted as $\gamma$). Physically, the GES reflects the vulnerability towards distorted state estimation that a single measurement outlier can induce. It is defined as the supremum of the IF norm:
\begin{equation} 
    \gamma = \sup_{r_k} \| \text{IF}(\psi, r_k) \|. 
\end{equation}

Since our objective is to compare the local robustness of the optimal LQLC estimator against the proposed LAH M-estimator toward an identical measurement anomaly, we may simplify the evaluation by neglecting the terms related to the number of measurements $n$ and the geometry matrix $\mathbf{H}$ in Equation \eqref{equ: def IF} and \eqref{equ: def IF M}. This simplification isolates the robustness characteristics of the selected loss kernel itself, allowing us to define the scalar residual GES ($\gamma_{\text{res}}$) as:

\begin{equation}
    \gamma_{\text{res}} = \frac{\sup_{r_k} |\psi(r_k)|}{\text{E}[\psi'(r)]}.
\end{equation}

Combined with Equation \eqref{aequ: E LQLC p} the residual GES for LQLC estimator is given by:
\begin{equation}
    \gamma_{\text{res}, \text{LQLC}} =  \frac{\sup_{r} |\psi_\text{LQLC}|}{\text{E}[\psi_\text{LQLC}']} = \frac{\sup_{r} \left| \frac{1}{s} \tanh\prt{\frac{r}{2s}} \right| }{\text{E}\left[\psi'_\text{LQLC}\right]} = \frac{1}{s\text{E}\left[\psi'_\text{LQLC}\right]}= \frac{1}{s\prt{\int_{-2s}^{2s} \frac{1}{2s^2} f(r) dr}}  .
\end{equation}

Besides, combined with Equation \eqref{aequ: E LAH p}, the residual GES for LAH estimator is derived by:
\begin{equation}
    \gamma_{\text{res}, \text{LAH}} =  \frac{\sup_{r} |\psi_\text{LAH}|}{\text{E}[\psi_\text{LAH}']}  = \frac{\sqrt{2}}{\sqrt{2}s} \cdot \frac{1}{\text{E}\left[\psi'_\text{LAH}\right]} = \frac{1}{s\text{E}\left[\psi'_\text{LAH}\right]}=\frac{1}{s\prt{\int_{-\infty}^{\infty} \left[ \frac{1}{2s^2}\text{sech}^2\left(\frac{r}{2s}\right) \right] f(r) dr}}  .
\end{equation}


\end{document}